\documentclass[iop]{emulateapj}

\usepackage[english]{babel}
\usepackage[T1]{fontenc}
\usepackage[utf8]{inputenc}

\usepackage[version=3]{mhchem}
\usepackage{graphicx}

\usepackage{units}
\usepackage{textcomp}
\usepackage{gensymb}

\slugcomment{To appear in The Astrophysical Journal Letters}

\shorttitle{Revisiting the scattering greenhouse effect of \ce{CO2} ice clouds}
\shortauthors{Kitzmann}

\begin{document}

\title{Revisiting the scattering greenhouse effect of \ce{CO2} ice clouds}

\author{D. Kitzmann\altaffilmark{1}}
\affil{Center for Space and Habitability, University of Bern, Sidlerstr. 5, 3012 Bern, Switzerland}

\begin{abstract}

Carbon dioxide ice clouds are thought to play an important role for cold terrestrial planets with thick \ce{CO2} dominated atmospheres. 
Various previous studies showed that a scattering greenhouse effect by carbon dioxide ice clouds could result in a massive warming of the planetary 
surface. However, all of these studies only employed simplified two-stream radiative transfer schemes to describe the anisotropic scattering.
Using accurate radiative transfer models with a general discrete ordinate method, this study revisits this important effect and shows that the 
positive climatic impact of carbon dioxide 
clouds was strongly overestimated in the past. The revised scattering greenhouse effect can have important implications for the early Mars, but also 
for planets like the early Earth or the position of the outer boundary of the habitable zone.

\end{abstract}


\section{Introduction}
Clouds are common in the atmospheres of Solar System planets and are likely ubiquitous in those of extrasolar planets as well. They 
affect every aspect of a planetary atmosphere, from radiative transfer, to atmospheric chemistry and dynamics, and they influence -- if not 
control -- aspects such as the surface temperature and, thus, the potential habitability of a terrestrial planet \citep{Marley2013cctp.book..367M}. 
Understanding the impact of clouds is thus instrumental for the study of planetary climates.

Clouds composed of carbon dioxide (\ce{CO2}) ice particles are important for cold, \ce{CO2}-rich atmospheres. This includes, for example, the present 
and early Mars \citep{Forget1997}, the early Earth \citep{Caldeira1992Natur.359..226C}, or terrestrial exoplanets at the outer boundary of the 
habitable 
zone \citep{Forget1997, Kasting1993}.
Provided that condensation nuclei are available, such clouds would form easily in a supersaturated atmosphere. 

Like for water clouds (e.g. \citet{Kitzmann2010A&A...511A..66K}), the presence of \ce{CO2} clouds will firstly result in an increase of the planetary 
albedo by scattering incident stellar radiation back to space which leads to a cooling effect (albedo effect). In contrast to liquid or solid 
water, \ce{CO2} ice is mostly transparent with respect to absorption in the infrared \citep{Hansen1997JGR, Hansen2005JGRE}.
Thus, as argued by \citet{Forget1997} or \citet{Kasting1993}, a classical greenhouse effect by absorption and re-emission of thermal radiation is 
unlikely to occur for a cloud composed of dry ice. On the other hand, \ce{CO2} ice particles can efficiently scatter thermal radiation back to the 
planetary surface, thereby exhibiting a scattering greenhouse effect. In their seminal study, Forget \& Pierrehumbert 
\cite{Forget1997} 
showed that this scattering greenhouse effect of \ce{CO2} ice clouds alone would have been efficient enough to allow for liquid water on the surface 
of the early Mars. 
This strong heating effect was also found in later studies  by e.g. \citet{Pierrehumbert1998JAtS}, \citet{Mischna2000Icar}, or 
\citet{Colaprete2003}. However, by using 
three-dimensional general circulation models, it has been suggested, that a limiting factor for the impact of 
\ce{CO2} clouds on early Mars might be the cloud coverage, which could be well below 100\% \citep{Forget2013Icar..222...81F}.

Especially, the outer boundary of the classical habitable zone (HZ) might also be influenced by the formation of \ce{CO2} ice 
clouds and their corresponding climatic impact. For the Sun, \citet{Forget1997} reported an outer HZ boundary of 2.4 au for a planet fully covered 
with \ce{CO2} clouds, which is far larger than the cloud-free value of 1.67 au \citet{Kasting1993}. 

The competing radiative cooling and heating effects are individually large, but partly balance each other. That means that small errors in the 
description of one of these radiative interactions can lead to large errors in a cloud's net effect.
So far, all previous studies on the climatic impact of \ce{CO2} clouds used a simplified treatment for the numerical solution of the radiative 
transfer equation to determine the atmospheric and surface temperatures. 
In particular, two-stream methods \citep{Toon1989JGR....9416287T} have exclusively been used. 
With only two angular directions available to determine the radiation field, these methods provide only limited means to describe e.g. the scattering 
phase function of the \ce{CO2} ice particles well enough to yield accurate results for describing radiative effects based on anisotropic scattering.

In a numerical radiative transfer study by \citet{Kitzmann2013A&A...557A...6K}, it was shown that these simplified two-stream methods underestimate 
the albedo effect at short wavelengths and overestimate the back-scattering of thermal radiation which forms the basis of the scattering greenhouse 
effect. However, this previous study did not use an atmospheric model and, thus, was unable to estimate the impact of a more accurate description of 
the scattering greenhouse effect on the surface temperature.

In this study, I will therefore re-asses the climatic impact of \ce{CO2} clouds using a sophisticated radiative transfer scheme. 
Since most of the studies related to the climatic impact of \ce{CO2} ice clouds have been done for early Mars, I will use the same modeling 
scenario to compare my atmospheric model calculations with the previously published results.

\section{Model description}

To investigate the impact of \ce{CO2} ice clouds, I use a one-dimensional radiative-convective climate model. The model incorporates 
a state-of-the-art radiative transfer treatment based on a discrete ordinate method \citep{Hamre2013AIPC.1531..923H} which is able to treat the 
anisotropic scattering by cloud particles accurately \citep{Kitzmann2013A&A...557A...6K}. 

The atmospheric model is stationary, i.e. it doesn't contain an explicit time dependence, and assumes hydrostatic equilibrium. About one hundred grid 
points are used to resolve the vertical extension of the atmosphere. The model currently considers \ce{N2}, \ce{CO2} and \ce{H2O}. Of those, only 
\ce{CO2} and \ce{H2O} are, however, used in this study.

The temperature profile is calculated from the requirement of radiative equilibrium by a time-stepping approach, as well as performing a convective 
adjustment, if necessary. The convective lapse rate is assumed to be adiabatic, taking into account the condensation of \ce{H2O}
and \ce{CO2}. A surface albedo of 0.215 based on measurements of present Mars is used \citep{Kieffer1977JGR....82.4249K}.
A more detailed model description will be presented in \citet{Kitzmann2016inprep}.

\subsection{Radiative transfer}

In contrast to many other atmospheric models for terrestrial exoplanets, the radiative transfer of this study is not separated into two different 
wavelength regimes. Instead, one single, consistent radiative transfer scheme within the wavelength range from \unit[0.1]{\micro m} to 
\unit[500]{\micro m} is used. 
At each distinct wavelength point, the plane-parallel radiative transfer equation
\begin{equation}
   \mu \frac{\mathrm{d} I_\lambda}{\mathrm{d}\tau_\lambda} = I_\lambda - S_{\lambda,\mathrm{*}}(\tau_\lambda) 
   \label{eq:rte}
\end{equation}
is solved, with the general source function
\begin{equation}
  \begin{split}
    S_{\lambda}(\tau_\lambda) = & S_{\lambda,\mathrm{*}}(\tau_\lambda) + (1 - \omega_\lambda) B_\lambda\\  
                              &   + \frac{\omega_\lambda}{2} \int_{-1}^{+1} p_\lambda(\mu,\mu')I_\lambda(\mu') \mathrm{d}\mu' \ ,
  \end{split}
\end{equation}
where $S_{\lambda,\mathrm{*}}$ is the contribution of the central star, $B_\lambda$ the Planck function, $\omega_\lambda$ the single scattering 
albedo, and $p_\lambda(\mu,\mu')$ the scattering phase function.

The scattering phase function is represented as an infinite series of Legendre polynomials \citep{Chandrasekhar1960ratr}
\begin{equation}
  p_\lambda(\mu,\mu') = \sum_{n=0}^{\infty} (2 n + 1) P_n(\mu) P_n(\mu') \chi_{\lambda,n}
\end{equation}
with the Legendre polynomials $P_n(\mu)$ and the phase function moments $\chi_{\lambda,n}$. In practice the series is truncated at a certain 
$n=N_\mathrm{max}$. For discrete ordinate methods the number of moments $N_\mathrm{max}$ is 
equal to the number of ordinates (streams) considered in the radiative transfer equation \citep{Chandrasekhar1960ratr}.

The equation of radiative transfer is solved by the discrete ordinate solver C-DISORT \citep{Hamre2013AIPC.1531..923H}. In contrast to 
two-stream methods, it yields the mathematically exact solution of the transfer equation for a given set of transport coefficients and phase 
functions, provided that enough computational streams are included.
For this study, eight streams are used for all model calculations. Doubling the number of streams has no impact on the resulting radiation 
fluxes and surface temperatures.

The wavelength-dependent absorption by atmospheric molecules and clouds is treated by the opacity sampling method, which is one of the standard 
methods employed in cool stellar atmospheres \citep{Sneden1976ApJ...204..281S}. In contrast to the $k$-distribution method with the correlated-$k$ 
assumption, opacity sampling still operates in the usual wavelength/wavenumber space \citep{Mihalas1978stat.book.....M}. 
This approach has the advantage, that absorption coefficients of all atmospheric constituents are 
fully additive. Essentially, opacity sampling can be regarded as a degraded line-by-line formalism. It is based 
on the fact that the wavelength integral of the radiation flux converges already well before 
all spectral lines are fully resolved. At each single wavelength, however, the solution with the opacity sampling method is identical to a 
corresponding line-by-line radiative transfer.

For this study, the distribution of the wavelength points at which the equation of radiative transfer is solved, is treated separately in three 
different wavelength regions. In the infrared, the points are sampled along the Planck black body curves for different temperatures. This method is 
adopted from \citet{Helling1998A&A} and guarantees an accurate treatment of the thermal radiation with a small number of wavelength points. In the 
visible and near infrared region, 10000 wavelength points are distributed equidistantly in wavenumbers. This region is most important for the 
temperature profile in the upper atmosphere where the temperatures are essentially determined by absorption of stellar radiation within 
well-separated lines. Beyond the visible wavelength region, about 100 points are used to cover the smooth Rayleigh scattering slope and resolving the 
stellar Lyman alpha emission line. In total about 12500 discrete wavelength points are used here. Tests by increasing the wavelength resolution show 
virtually no change in the wavelength-integrated flux. In fact, the high number of integration points in the near infrared region could also be 
decreased if one is not interested in the temperatures near the top-of-the atmosphere.

Absorption cross-sections for \ce{CO2} and \ce{H2O} are calculated as described in \citet{Wordsworth2013ApJ...778..154W}, using the HITRAN 2012 
database \citep{Rothman2013JQSRT.130....4R}. For these calculations, the open source \textit{Kspectrum} code (version 1.2.0) is used. In case of 
\ce{CO2}, the sub-Lorentzian line profiles of \citet{Perrin1989JQSRT..42..311P} are employed, while the contribution of collision induced absorption 
is 
taken from \citet{Baranov2003JMoSp.218..260B}. The continuum absorption of \ce{H2O} is derived from the MT-CKD description 
\citep{Mlawer2012RSPTA.370.2520M}. Rayleigh scattering is considered for \ce{CO2} and \ce{H2O} \citep{vonParis2010A&A...522A..23V}.

\subsection{Cloud description}

The atmospheric model also takes the radiative effect of clouds directly into account. Following \citet{Forget1997}, the size distribution 
of the \ce{CO2} ice particles is described by a modified gamma distribution
\begin{equation}
  f(a) = \frac{\left( a_\mathrm{eff} \nu \right)^{2-1/\nu}}{\Gamma\left(\frac{1-2\nu}{\nu}\right)} a^{\left(\frac{1}{\nu}-3\right)} \mathrm 
e^{-\frac{a}{a_\mathrm{eff}\nu}} \,
  \label{eq:gamma_distribution}
\end{equation}
described by the effective radius $a_\mathrm{eff}$ and an effective variance of 0.1 \citep{Forget1997}. 

\begin{figure}
        \includegraphics[width=0.45\textwidth]{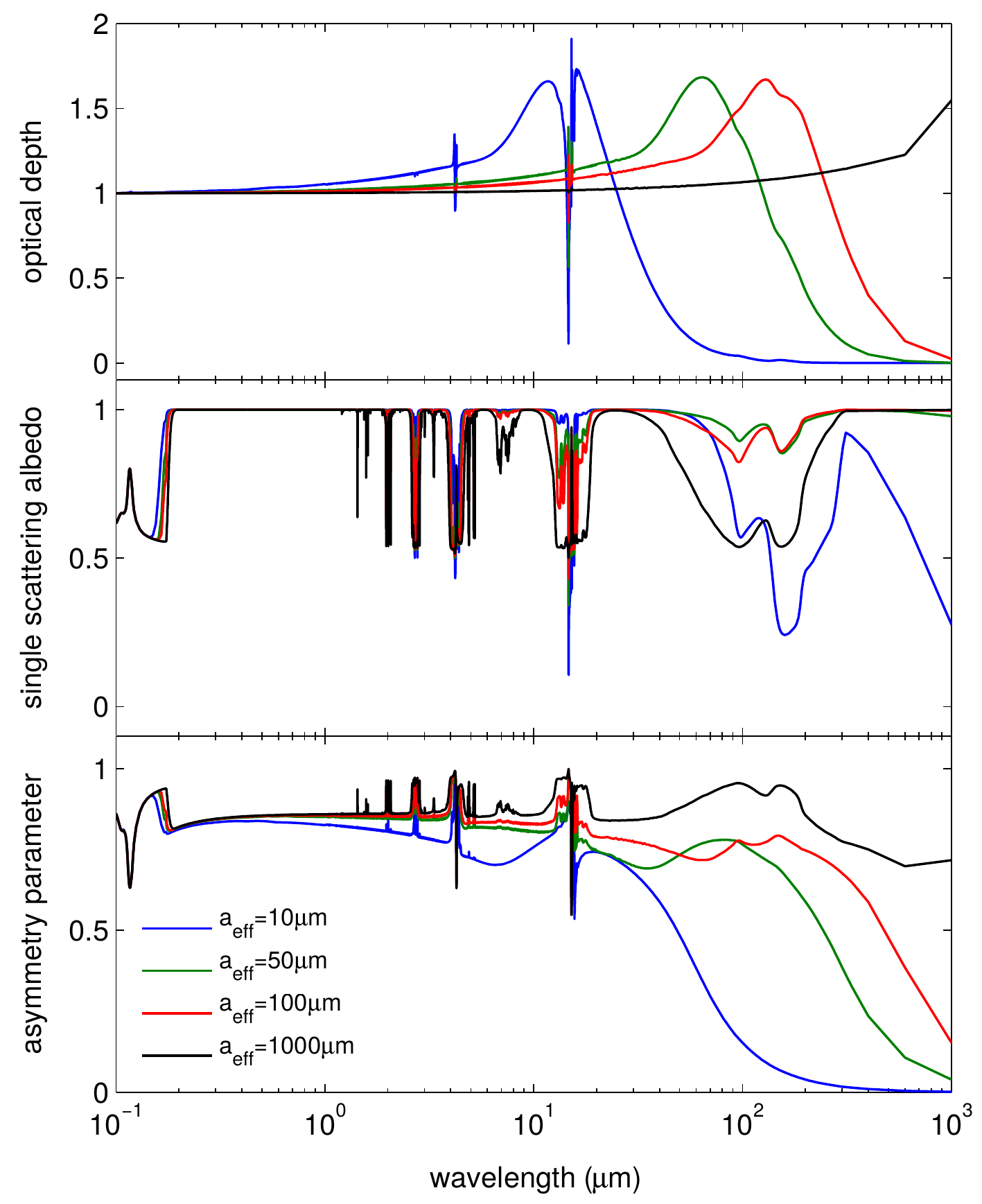}
	\caption{\textbf{Optical properties of CO$_2$ ice particles.} Results are shown for four different size distributions: 
		$a_{\mathrm{eff}}=10 \ \mathrm{\mu m}$ (\textit{blue line}), $a_{\mathrm{eff}}=50 \ \mathrm{\mu m}$ (\textit{green line}), 
$a_{\mathrm{eff}}=100.0 \ 
		\mathrm{\mu m}$ (\textit{red line}), $a_{\mathrm{eff}}=1000 \ \mathrm{\mu m}$ (\textit{black line}). \textit{Upper diagram}: optical 
depth (for $\tau = 1$), \textit{middle diagram}: single scattering albedo, \textit{lower diagram}: asymmetry parameter.}
	\label{fig:optical_properties}
\end{figure}

Assuming spherical particles, the optical properties are calculated via Mie theory, using the refractive index of \ce{CO2} ice from 
\citet{Hansen1997JGR,Hansen2005JGRE}. 
The resulting optical properties are shown in Fig. \ref{fig:optical_properties} for some selected values of the effective radius and an optical depth 
of one. The optical depth $\tau$ of the clouds refers to the particular wavelength of $\lambda = 0.1 \ \mathrm{\mu m}$ 
throughout this study. 

The Henyey-Greenstein function \citep{Henyey1941ApJ....93...70H} is used to approximate the scattering phase function. Although 
it lacks the complicated 
structure and detailed features of the full Mie phase function, the Henyey-Greenstein function preserves its average quantities, such as the 
asymmetry parameter $g$. The Henyey-Greenstein phase function is usually a very good replacement of the Mie phase function, especially at 
higher optical depths or if one is only interested in angular averaged quantities, such as the radiation flux  \citep{vandeHulst1968JCoPh...3..291V, 
Hansen1969JAtS...26..478H}.

\section{Effect of \ce{CO2} clouds in the early Martian atmosphere}

To compare the climatic impact of \ce{CO2} ice clouds with previous model studies, I use the same model set-up as in \citet{Forget1997}, who studied 
the influence of \ce{CO2} clouds in the atmosphere of the early Mars.

Thus, following \citet{Forget1997} or \citet{Mischna2000Icar}, a \ce{CO2} surface pressure of 2 bar is used on a planet with the radius 
and mass of Mars. To simulate the conditions for early Mars with a less luminous Sun, the incident stellar flux is set to 75\% of the present-day 
Martian Solar irradiance. A high-resolution spectrum of the present-day Sun from \citet{Gueymard2004} is used for the spectral energy distribution of 
the incident stellar flux.

\begin{figure}
	\includegraphics[width=0.45\textwidth]{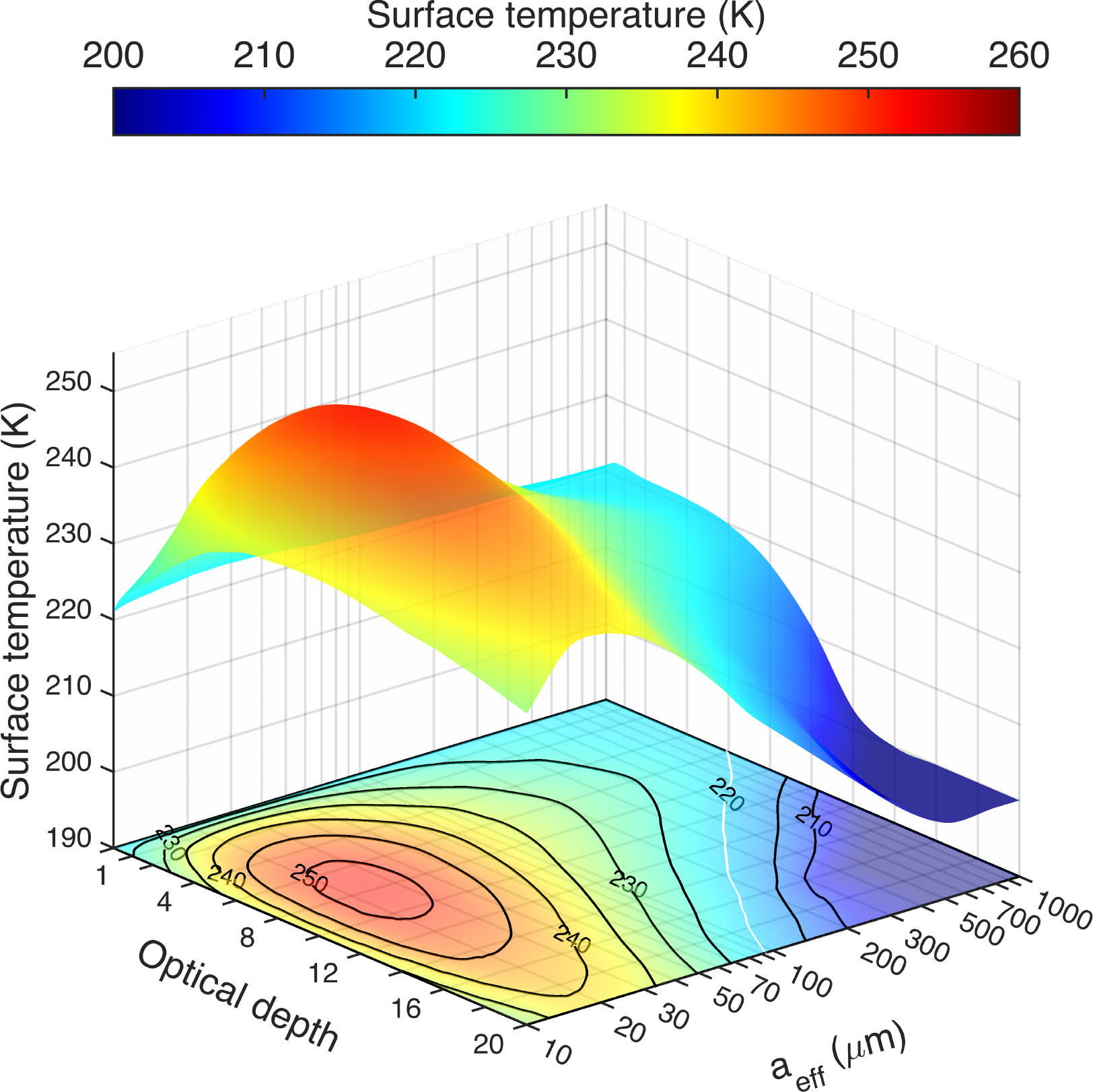}
	\caption{\textbf{Impact of \ce{CO2} ice clouds on the surface temperature.} Surface temperatures are shown as a function of 
optical depth (at a wavelength of $\lambda = 0.1 \ \mathrm{\mu m}$) and effective particle size of the \ce{CO2} ice clouds. The black contour 
lines are given in steps of 5 K. The white contour line indicates the clear-sky case.}
	\label{fig:surface_temp_2d}
\end{figure}

Because a supersaturated \ce{CO2}-rich atmosphere would provide a huge amount of condensible material, dry ice 
particles would grow rapidly, thereby potentially reaching large particle sizes under these conditions. The only detailed microphysical study for 
particle formation on early Mars obtained mean particle sizes of the order of \unit[1000]{\micro m} \citep{Colaprete2003}. 
Thus, to cover this very wide parameter range, the effective particle sizes are varied between \unit[10]{\micro m} and \unit[1000]{\micro m} in the 
following. A total cloud coverage is assumed to obtain an upper limit for the clouds' climatic impact. The cloud layer is placed into the 
supersaturated region of the cloud-free atmosphere around a pressure of 0.1 bar. This corresponds roughly to the 
position of the cloud layer in \citet{Forget1997} and \citet{Mischna2000Icar}.

Since the publication of \citet{Forget1997} or \citet{Mischna2000Icar}, several updates to the molecular line lists and description of the molecule's 
continuum absorption were introduced. While the greenhouse effect of \ce{CO2} became less effective \citep{Wordsworth2010Icar..210..992W}, the  
absorptivity of \ce{H2O} increased, especially due to changes in the continuum absorption \citep{Mlawer2012RSPTA.370.2520M} and the line 
parameters \citep{Rothman2013JQSRT.130....4R}. Therefore, to limit the impact of the different molecular line lists on the comparison, only \ce{CO2} 
is considered as an atmospheric gas in the following. This corresponds to the ``dry case'' in \citet{Forget1997}. 
 
In the clear-sky case, the surface temperature is about 220 K and thus roughly 8 K smaller than reported by \citet{Forget1997} and 
\citet{Mischna2000Icar}. This is caused by a revised collision induced absorption (CIA) of the \ce{CO2} gas, yielding a smaller greenhouse 
effect \citep{Wordsworth2010Icar..210..992W}. 

The resulting surface temperatures for the cloudy cases are shown in Fig. \ref{fig:surface_temp_2d} as a function of the 
cloud particle sizes and optical depth. In total, several hundreds of individual model calculations have been performed to obtain the results in Fig. 
\ref{fig:surface_temp_2d}.

In agreement with previous studies \citep{Forget1997, Pierrehumbert1998JAtS, Mischna2000Icar, Colaprete2003}, \ce{CO2} particles with effective radii 
larger than about \unit[10]{\micro m} can result in a net greenhouse effect. 
An efficient greenhouse effect, however, is only possible within a certain particle size range and for medium values of the optical depth. 
Particles larger than roughly \unit[500]{\micro m} are more or less neutral or - at high optical depths - even exhibit a net cooling effect. 
These particles are too large to feature an efficient back-scattering of the upwelling thermal radiation. Their large asymmetry parameters at thermal 
infrared wavelengths results in thermal radiation being predominantly scattered in the upward direction and, thus, away from the surface. 
In order to obtain the highest possible greenhouse effect, the particle sizes must be comparable to the wavelength of the
atmospheric thermal radiation below the cloud layer (cf. also \citet{Kitzmann2013A&A...557A...6K}). This is only the case for $a_{\mathrm{eff}} 
\approx \unit[25]{\micro m}$ which therefore provides the largest net greenhouse effect.

In no case studied here, the freezing point of water is reached - quite in contrast to previous studies where 
surface temperatures in excess of \unit[300]{K} had been found for the same scenario \citep{Forget1997}.
The highest surface temperatures obtained for a pure 2 bar \ce{CO2} atmosphere are about \unit[252]{K}. 
Even considering the \unit[8]{K} difference due to the 
revised \ce{CO2} CIA, the differences in the resulting surfaces temperatures between the different radiative transfer approaches are astonishingly 
large. However, it should be noted, that the lower collision induced absorption of the \ce{CO2} molecules might also influence the 
resulting scattering greenhouse effect of the ice particles to a certain extend.

\begin{figure}
	\includegraphics[width=0.45\textwidth]{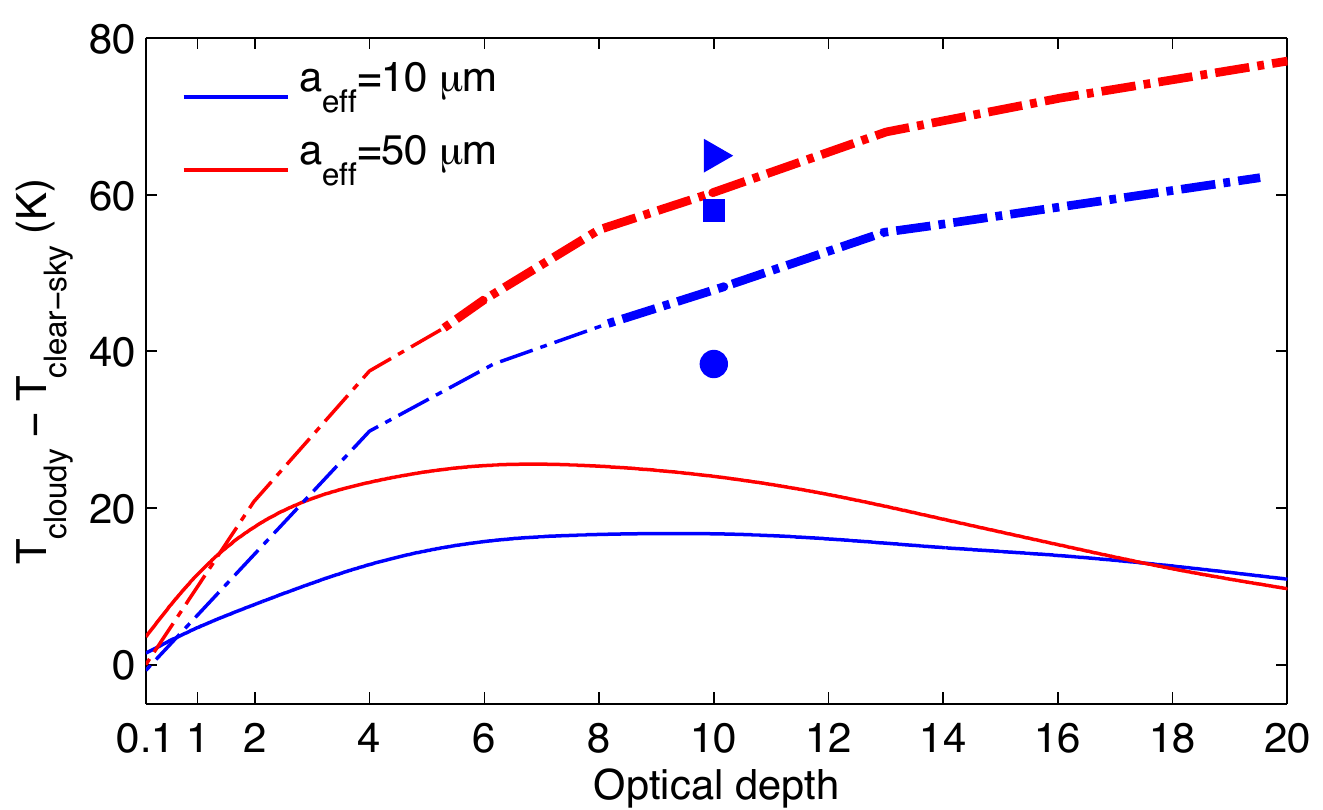}
	\caption{\textbf{Comparison of the impact of \ce{CO2} clouds on the surface temperature with previous studies.} The diagram shows the changes of the 
		surface temperature in the presence of \ce{CO2} ice clouds ($T_\mathrm{cloudy}$) with respect to the clear-sky case ($T_\mathrm{clear-sky}$), 
		compared to the results of \citet{Forget1997} and \citet{Mischna2000Icar}. Thicker lines refer to surface temperatures 
                above the freezing point of water.
		The results are shown for two effective radii: $a_{\mathrm{eff}} =$ \unit[10]{\micro m} (blue) and $a_{\mathrm{eff}} =$ 
                \unit[50]{\micro m} (red). Solid 
		lines denote the results from this study, dashed-dotted lines the dry atmosphere results from \citet{Forget1997} for comparison.
		The single three blue markers compare results for 
		an atmosphere fully saturated with water (circle: this study, square: \citet{Mischna2000Icar}, triangle: \citet{Forget1997}).}
	\label{fig:contour_sizes}
\end{figure}

A direct comparison of the climatic impact with the previous studies of \citet{Forget1997} and \citet{Mischna2000Icar} is shown in Fig.  
\ref{fig:contour_sizes}. Note, that \citet{Forget1997} only used two effective radii (10 and \unit[50]{\micro m}), while \citet{Mischna2000Icar} is 
limited to a single value of \unit[10]{\micro m} and studies only atmospheres saturated with water vapor.

\begin{figure}
  \centering
  \includegraphics[width=0.35\textwidth]{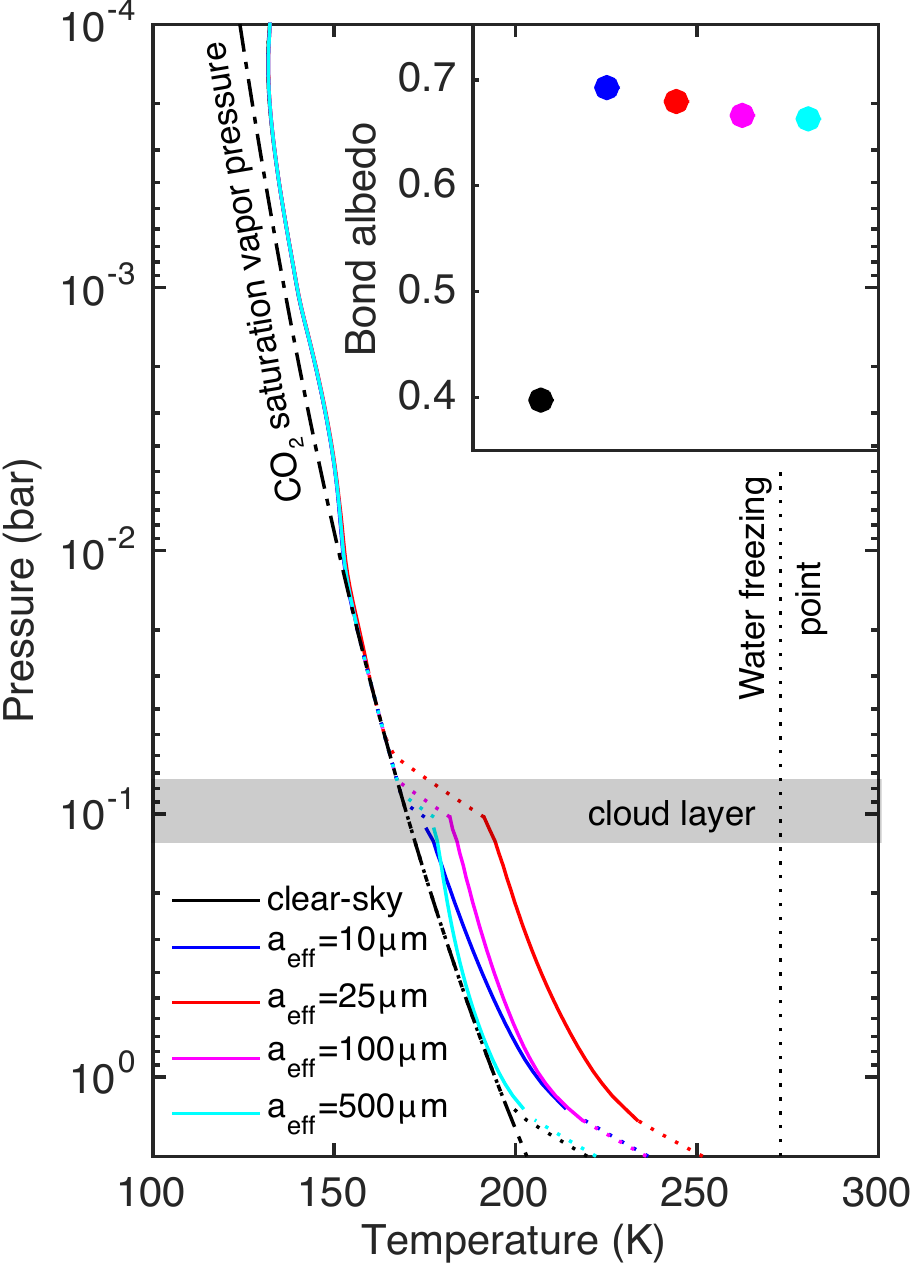}
  \caption{\textbf{Atmospheric temperature-pressure profiles for the dry case influenced by \ce{CO2} clouds.} The resulting temperature profiles 
           are shown for the clear-sky case (black line) and for several values of the effective radius of the particle size distribution 
           (see legend). The optical depth is 8 in each case.
           The effect of \ce{CO2} clouds on the energy transport is marked by different line-styles. Solid lines indicate a temperature profile 
           in radiative	equilibrium, dashed lines dry convective regions, and dashed-dotted a moist \ce{CO2} adiabatic lapse rate. The 
           vertical dotted line denotes the freezing point of water. The dashed-dotted line marks the saturation vapor pressure curve of \ce{CO2}.
           The position of the cloud layer is denoted by the grey-shaded area.
           The inset plot shows the planetary Bond albedos affected by the presence of the \ce{CO2} ice clouds. Colours refer to the same 
           effective particle sizes mentioned above.}
  \label{fig:temperture_profiles}
\end{figure}

The results in Fig. \ref{fig:contour_sizes} clearly suggest that the climatic impact of \ce{CO2} clouds was strongly overestimated in the past. 
The cloud-induced temperature changes determined here are several tens of Kelvin smaller in most cases. 
For an optical depth of 10 and an effective radius of \unit[10]{\micro m} also the results for a water-saturated atmosphere are compared. Again, 
the temperature increases due to the scattering greenhouse effect are between 20 and 30 K smaller than those found in \citet{Forget1997} or 
\citet{Mischna2000Icar} for this particular case. 
Additionally, \citet{Colaprete2003} also claimed a strong greenhouse effect for particle sizes of several hundred \unit{\micro m}, 
which, as already shown in Fig. \ref{fig:surface_temp_2d}, is clearly not the case.

The presence of \ce{CO2} clouds also has a profound impact on the temperature profile, as shown in Fig. \ref{fig:temperture_profiles}. 
Due to their scattering greenhouse effect, they strongly increase the atmospheric temperature \textit{locally}, just below the cloud base. This 
inhibits convection and creates a temperature profile where the usual fully convective troposphere is changed into a lower convective region near the 
surface and a second one above the cloud layer. 
Both convective regions are separated by a temperature profile determined by radiative equilibrium (see also \citet{Mischna2000Icar}). The local 
heating due to the cloud particles is in fact strong enough to cause their evaporation. Thus, such a cloud layer would be unable to persist in 
stationary equilibrium \citep{Colaprete2003}. Placing the cloud layer even higher up in the atmosphere would also result in the strong local heating.

Overall, the Bond albedos in the cloudy cases are somewhat higher than the ones of the reported by the previous studies. This is caused by the 
underestimation of the cloud's albedo effect by the two-stream radiative transfer schemes \citep{Kitzmann2013A&A...557A...6K}.

\section{Summary}

In this study, the potential impact of scattering greenhouse effect of \ce{CO2} ice clouds on the surface temperature is revisited by using an 
atmospheric model with an accurate radiative transfer method. By comparison with previous model studies on the early Mars, the results suggest 
that the potential heating effect was strongly overestimated in the past.

Based on the results presented here, it is, therefore, strongly recommended that atmospheric models which include the climatic effect of \ce{CO2} ice 
clouds should employ more suitable radiative transfer schemes. Additionally, previous model calculations involving the effect of \ce{CO2} ice clouds, 
such as for the position of the outer boundary of the habitable zone or the atmosphere of early Mars for example, clearly require to be revisited. 
This study also emphasizes the importance of using detailed radiative schemes when studying phenomena based on anisotropic scattering.

While the results for the scattering greenhouse effect found in this study are by far smaller than those reported by previous studies, this doesn't 
mean that \ce{CO2} ice clouds are overall ineffective in warming planetary atmospheres. On the contrary, they can still provide an important 
contribution to the greenhouse effect in a certain parameter range. Thus, the original idea brought up by Forget \& Pierrehumbert in their pilot 
study about a net scattering greenhouse effect still prevails, albeit much smaller than originally anticipated.

\acknowledgments

D.K. would like to thank Y. Alibert, K. Heng, J. Lyons, and J. Unterhinninghofen for their suggestions and comments and especially B. 
Patzer for the fruitful discussions. D.K also gratefully acknowledges the support of the Center for Space and Habitability of the University of Bern.

\bibliography{references}
\bibliographystyle{apj}

\end{document}